\documentclass{article}

\usepackage{amsmath,graphicx}
\usepackage{spconf}
\usepackage[comma,sort&compress,numbers]{natbib}

\usepackage{color,soul}

\definecolor{gray}{rgb}{1,0.6,0.5}

\usepackage[font=footnotesize]{caption}
\usepackage{booktabs}
\usepackage{eqparbox}
\usepackage{multirow}
\usepackage{algorithm,algorithmic}
\usepackage{flushend}

\newcommand{\CASE}[1]{\STATE \textbf{case} #1\textbf{:} \begin{ALC@g}}
\newcommand{\ENDCASE}{\end{ALC@g}}

\newcommand{\DEFAULT}{\STATE \textbf{default:} \begin{ALC@g}}
\newcommand{\ENDDEFAULT}{\end{ALC@g}}
\newcommand{\DEFAULTLINE}[1]{\STATE \textbf{default:} }

\newcommand*{\fnref}[1]{\textsuperscript{\ref{#1}}}

\graphicspath{{figures/}}

\title{An adaptive Lagrange multiplier determination method for rate-distortion optimisation in hybrid video codecs}
\name{Fan Zhang and David R. Bull \thanks{*This work was funded by UK EPSRC grant EP/J019291/1.}}
\address{Department of Electrical and Electronic Engineering, University of Bristol, Bristol, BS8 1UB, UK.\\
{\{Fan.Zhang, Dave.Bull\}@bristol.ac.uk}}
\begin{document}

\maketitle
\begin{abstract}
This paper describes an adaptive Lagrange multiplier determination method for rate-quality optimisation in video compression. Inspired by the experimental results of a Lagrange multiplier selection test, the presented approach adaptively estimates the optimum Lagrange multiplier for different video content, based on distortion statistics of recently encoded frames. The proposed algorithm has been fully integrated into both the H.264 and HEVC reference codecs, and is used in rate-distortion optimisation for encoding B frames. The results show promising (up to 11\% on the sequences tested) overall bitrate savings, for a minimal increase in complexity, on various types of test content based on Bjontegaard delta measurements.

\end{abstract}
\section{Introduction}
\label{sec:intro}

It is currently a very exciting and challenging time for video compression - the predicted growth in demand for bandwidth by video applications is probably greater now than it has ever been \cite{r:cisco}. While advances in network and physical layer technologies will no doubt contribute to the solution, the role of video compression is also of key importance. 

The last two decades have seen impressive improvements in the rate-distortion performance of video codecs. The most recent standardised codec, High Efficiency Video Coding (HEVC) \cite{j:HEVC}, offers a 50\% reduction in bit rate compared to H.264/AVC \cite{j:Sullivan1}. This is due to the addition of various novel features including larger block structures, additional intra-prediction directions, adaptive motion vector prediction, enhanced entropy coding and new deblocking filters. It is however noted that the rate-distortion optimisation (RDO) approach - a key component in video compression - remains almost the same in HEVC as in previous standard codecs. 

The current RDO model employed in HEVC and H.264/AVC was initially presented by Sullivan and Wiegand \cite{j:Sullivan2}, in which a Lagrange multiplier method is efficiently employed to determine the optimum compression modes for various coding unit sizes, as well as to search for the optimal motion vectors in motion estimation. This model is based on entropy-constrained high-rate approximation \cite{b:jayant,j:Gish}, and a Lagrange multiplier determination experiment \cite{c:Wiegand}. Although this approach has been widely employed, due to its simplicity and efficiency, it has however been found to have shortcomings in low bit-rate applications \cite{j:Li} which are not applicable under the high-rate assumption \cite{b:jayant,j:Gish}. Improved Lagrange multiplier selection methods were thus proposed, e.g. the $\rho$-domain based algorithm \cite{c:Chen}, and the Gaussian and Laplace distribution based methods \cite{j:Zhao,j:Li}. 

This paper investigates the optimality of the current Lagrange multiplier determination method in applications using constant quantisation parameters, and reports a consistent correlation between the distortion statistics and optimum Lagrange multiplier values. This has been exploited to adaptively modify Lagrange multipliers for RDO in H.264 and HEVC, providing consistent bitrate savings over both anchor algorithms with only a small increase in complexity.

The remainder of the paper is organised as follows. Section \ref{sec:experiment} describes the experiment on Lagrange multiplier selection and analyses the results. In Section \ref{sec:algorithm}, the adaptive Lagrange multiplier selection method is presented in detail, while the compression results of the proposed method are given in Section \ref{sec:results}. The conclusions are then  presented in Section \ref{sec:conclusion}. 

\section{The experiment on Lagrange multiplier selection}
\label{sec:experiment}

The rate-quality optimisation algorithm \cite{j:Sullivan2,c:Wiegand} employed in HEVC and H.264/AVC is based on a sufficiently high rate assumption, and the optimum coding parameters $\mathbf{p}_\mathrm{opt}$ are selected in order to minimise the cost function of distortion $D$ and rate $R$ \cite{b:Bull}:

\vspace{-10pt}
\small
\begin{equation}
\mathbf{p}_\mathrm{opt} = \underset{\mathbf{p}}{\mathrm{argmin}}\{D(\mathbf{p}) + \lambda R(\mathbf{p})\},
\end{equation}
\normalsize
where $\mathbf{p}$ is the vector of coding parameters, and $\lambda$ is the Lagrange multiplier. A sum of squared difference (SSD) distortion measure is normally employed and this leads to the following equations for Lagrange multiplier calculation for H.264/AVC I, P, and B frames\footnote{Here we follow the same frame type definition as in H.264/AVC. It is also noted that both P and B frames are defined as B frames in HEVC. In this paper, unless otherwise specified, we employ B frames where both temporally previous and forward frames are used as reference, and use P frames where only previous frames are used as reference.}.

\vspace{-10pt}
\small
\begin{equation}
\left\{ 
\begin{array}{ll}
\lambda_{\mathrm{H.264(I)}} &= 0.57 \cdot 2^\frac{\mathrm{QP}-12}{3}\\
\lambda_{\mathrm{H.264(P)}} &= 0.85 \cdot 2^\frac{\mathrm{QP}-12}{3}\\
\lambda_{\mathrm{H.264(B)}} &= 0.68 \cdot \mathrm{max}(2,\mathrm{min}(4,\frac{\mathrm{QP}-12}{6})) \cdot 2^\frac{\mathrm{QP}-12}{3}\\
\end{array}
\right.
\label{eq:lambdaH264}
\end{equation} 
\normalsize
This model is also used in HEVC with a minor revision.

\vspace{-10pt}
\small
\begin{equation}
\left\{ 
\begin{array}{ll}
\lambda_{\mathrm{HEVC(I)}} &= (1-\mathrm{max}(0,\mathrm{min}(0.5,0.05 N_B)))\cdot 0.57 \cdot 2^\frac{\mathrm{QP}-12}{3}\\
\lambda_{\mathrm{HEVC(P)}} &=  p \cdot 2^\frac{\mathrm{QP}-12}{3}\\
\lambda_{\mathrm{HEVC(B)}} &=  p \cdot \mathrm{max}(2,\mathrm{min}(4,\frac{\mathrm{QP}-12}{6})) \cdot 2^\frac{\mathrm{QP}-12}{3}\\
\end{array}
\right.
\label{eq:lambdaHEVC}
\end{equation}
\normalsize 
Here $N_B$ is the number of B frames in one group of pictures (GOP), and $p$ is a parameter, configurable before encoding, with a default value 0.5. The latter can also be assigned different values for different frame postions in a GOP.

It is noted that Lagrange multipliers in both H.264/AVC and HEVC mainly depend on QP values, and are not directly related to the characteristics of the input video signal. Moreover, this model is based on the experiment in \cite{c:Wiegand} that used an H.263 codec (without B frames) and four QCIF ($176\times144$) test sequences. With recent advances in video compression, especially with bi-directional prediction and intra coding in inter frames, the RD characteristic of codecs with B frames is likely to be quite different. 

In order to investigate the optimum Lagrange multiplier determination for B frame encoding, an Lagrange multiplier selection experiment was applied to both H.264 and HEVC reference codecs, in which the RD performance using original values $\lambda_\mathrm{orig}$ in both codecs - according to (\ref{eq:lambdaH264}) and (\ref{eq:lambdaHEVC}) -  are compared with the performance when different test Lagrange multipliers are used for B frames\footnote{We only changed the $\lambda$ values for B frames, and those for I and P frames were kept the same.}. The test values $\lambda_\mathrm{test}$ used are given in (\ref{eq:test}).
\begin{equation}
\lambda_\mathrm{test} = k \cdot \lambda_\mathrm{orig}.
\label{eq:test}
\end{equation}    
where the range of k is from $0.2$ to $5$, and this parameter was constant during encoding. 

\begin{table}[htbp]
\centering
\footnotesize
\caption{Test videos used in the Lagrange multiplier selection experiment.}
\begin{tabular}{c|l|l}
\toprule
Class  & Sequence			& Source\\
\midrule
A					& 1. Carpet, 2. Miss-America, 3. Picture & BVI \& Standard\\
 														\midrule
B				  & 4. Flag, 5. Spring, 6. Water  & DynTex  \\
 														\midrule
C				  & 7. Football, 8. Flower   9. Mobile  &  Standard \& DynTex\\
\bottomrule			
\end{tabular}
\label{tab:testvideos}
\end{table}

Primary coding settings used in this experiment include: JM 15.1 for H.264 and HM 14.0 for HEVC; uniform QP used for all frames - QP 27, 32, 37 and 42 tested for H.264 and for HEVC; three B frames used for both codecs (GOP length of 4); main profile and non-hierarchical B frames are used for H.264; main profile and hierarchical B frames are used for HEVC; SSD is used as distortion metric for RDO in both codecs. 

Nine video sequences with 101 frames (25 GOPs) at CIF resolution (YUV 4:2:0) are employed here for this experiment including 3 clips with slow movement (Class A), 3 dynamic texture sequences (Class B), and 3 videos with mixed content (Class C). These sequences are either standard test sequences, from the public databases DynTex \cite{j:dyntex} or from the BVI texture database \cite{w:bviTex}. These sequences are all available online\footnote{http://data.bris.ac.uk/data/dataset/1oio9r9fm8byb1ia7mtv0sct0t\label{ftn:webdata}}. Table \ref{tab:testvideos} provides a brief summary of the sequences used.

\begin{figure}[ht]
\scriptsize
\centering
\begin{minipage}[c]{0.49\linewidth}
  \centering
  \centerline{\includegraphics[width=4.6cm]{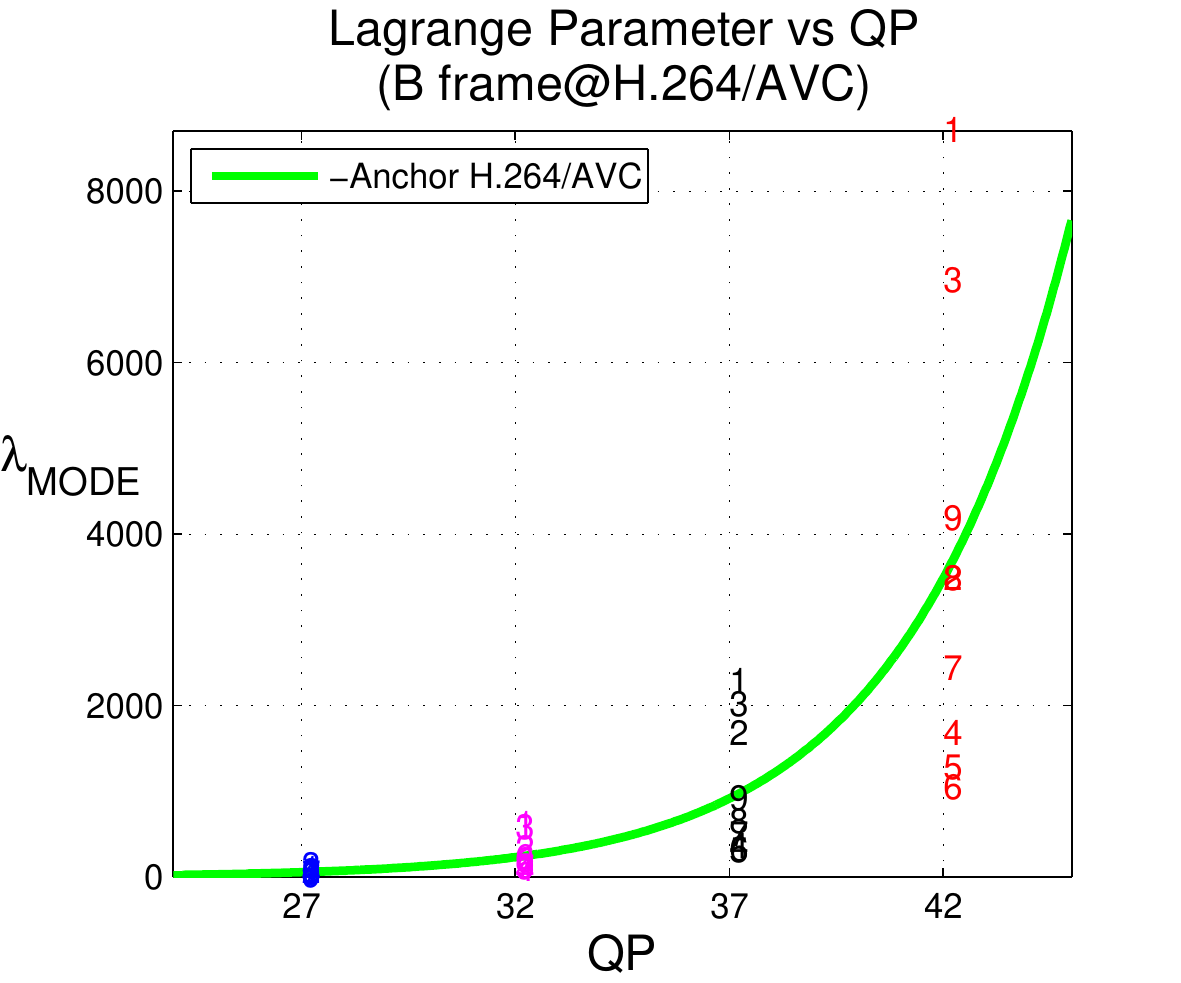}}
\end{minipage} 
\begin{minipage}[c]{0.49\linewidth}
  \centering
  \centerline{\includegraphics[width=4.6cm]{best_lambda_hevc.pdf}}
\end{minipage} 
\begin{minipage}[c]{0.49\linewidth}
  \centering
  \centerline{\includegraphics[width=4.6cm]{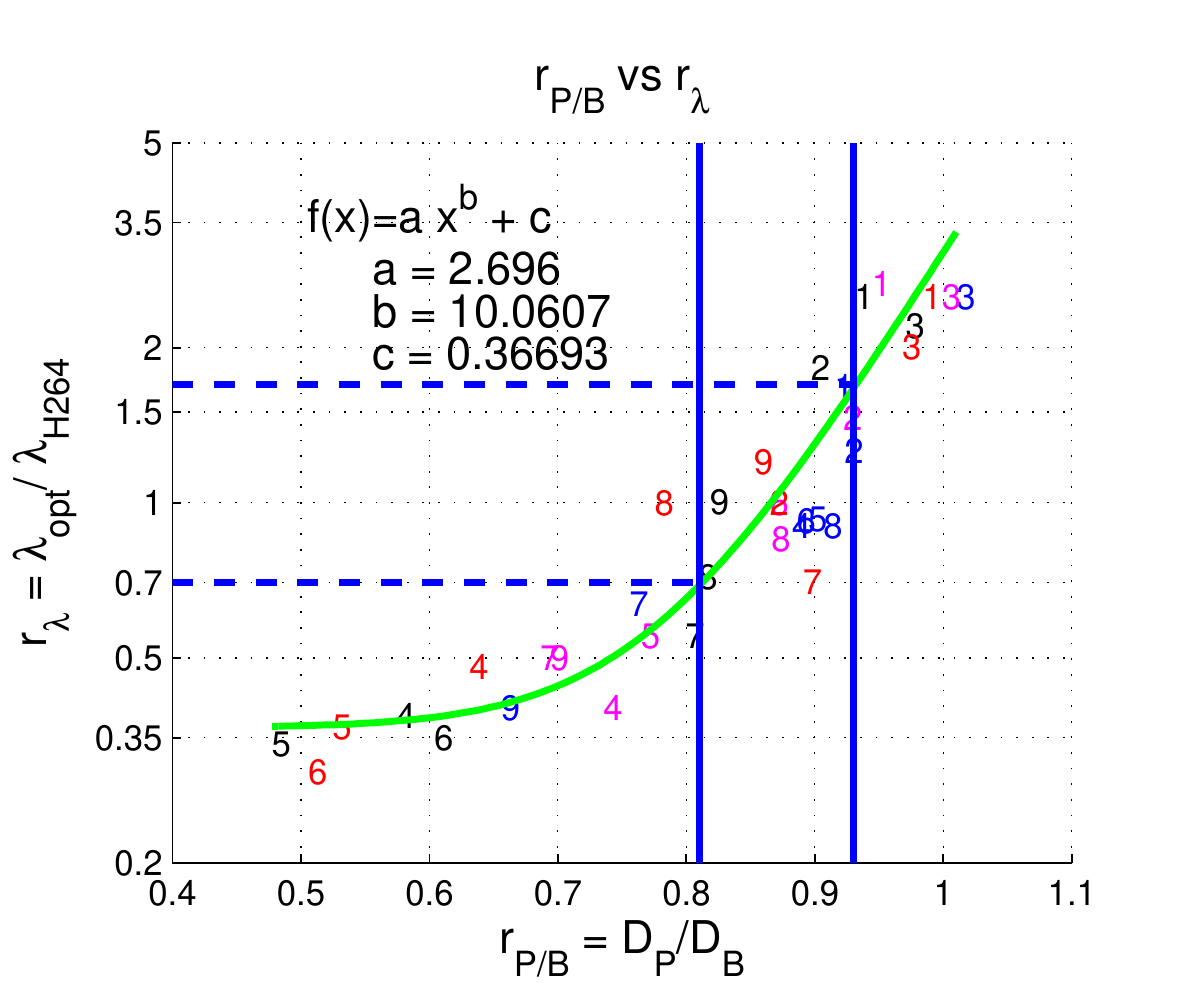}}
\end{minipage} 
\begin{minipage}[c]{0.49\linewidth}
  \centering
  \centerline{\includegraphics[width=4.6cm]{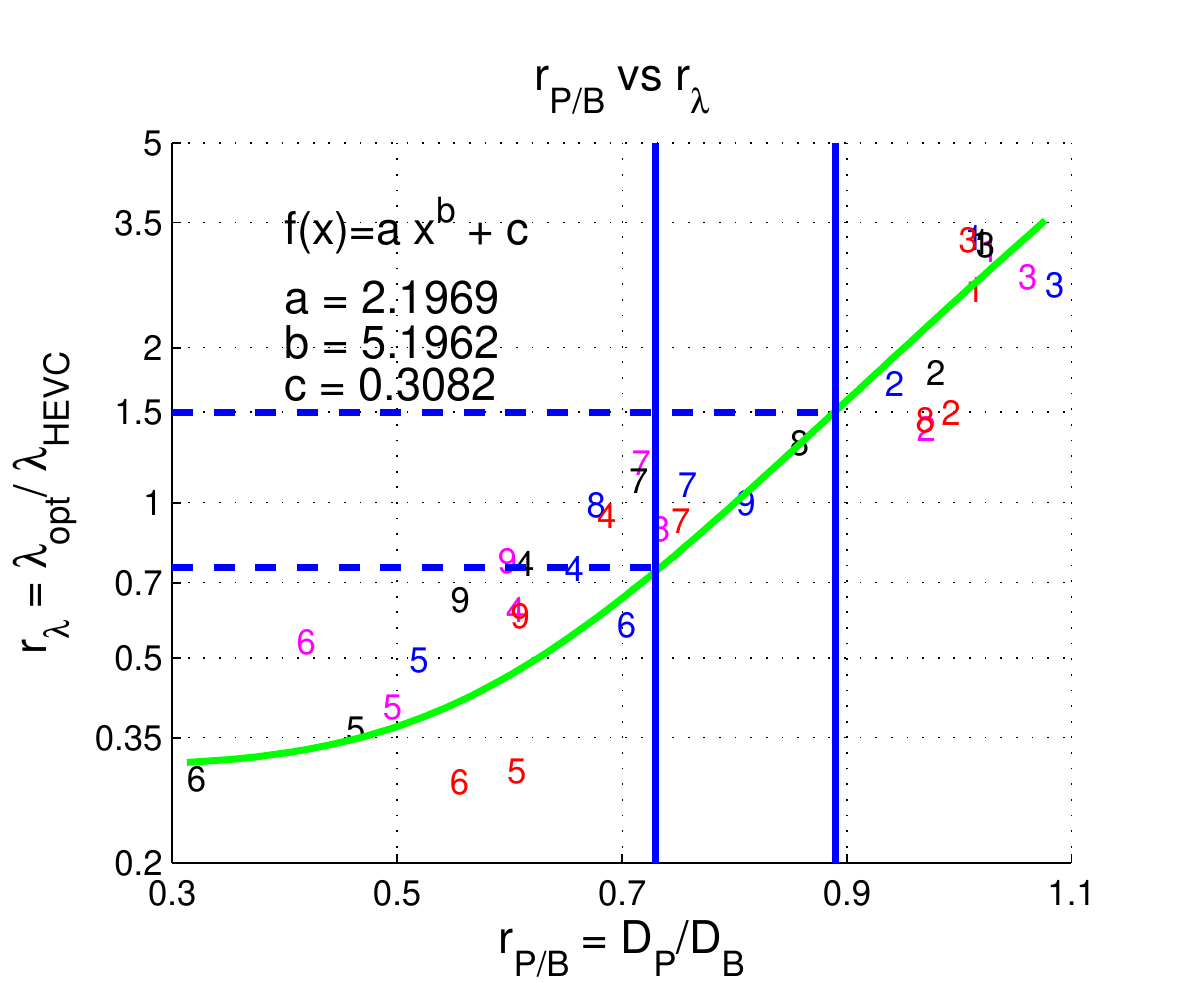}}
\end{minipage} 
\begin{minipage}[c]{0.49\linewidth}
  \centering
  \centerline{\includegraphics[width=4.6cm]{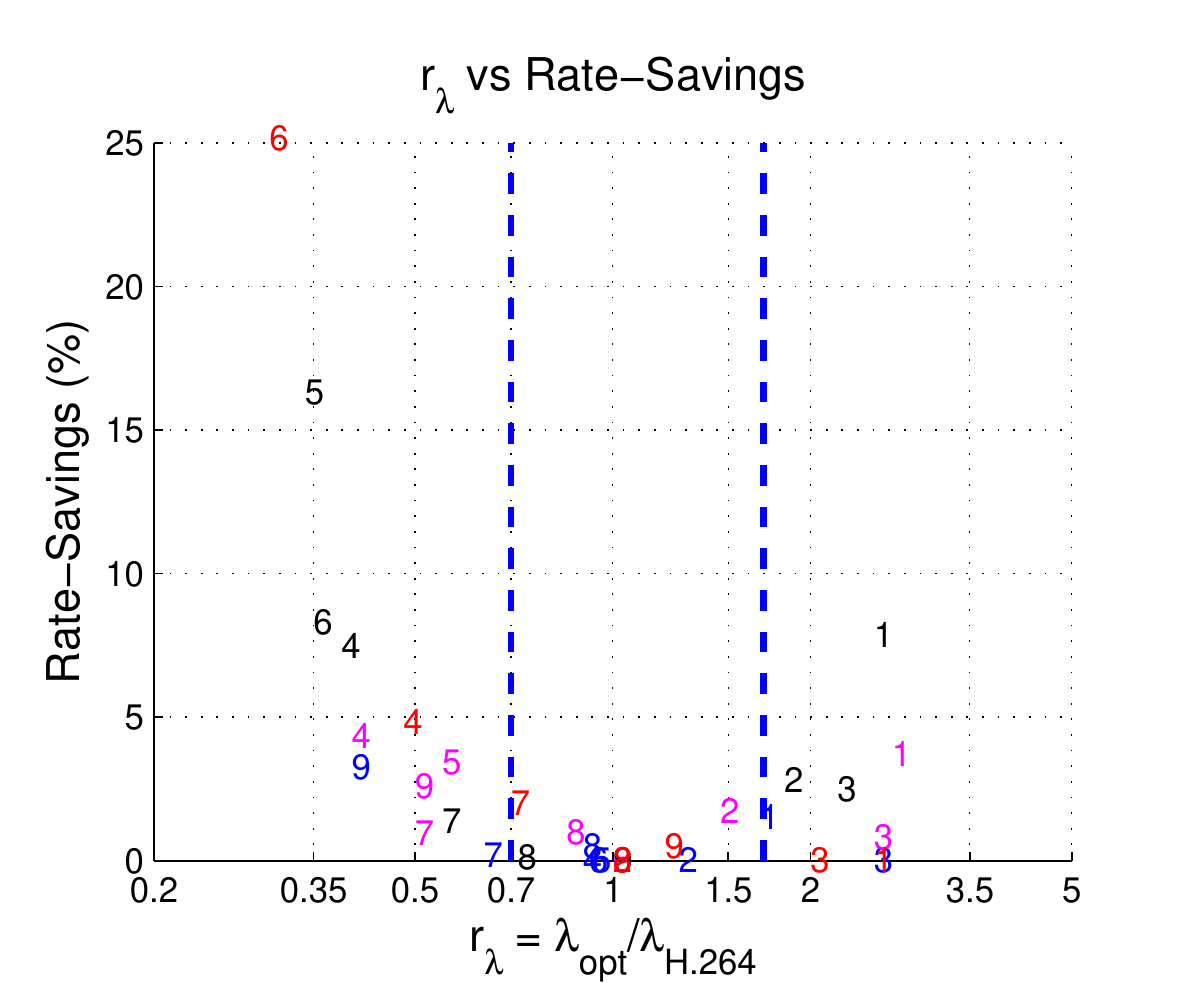}}
\end{minipage} 
\begin{minipage}[c]{0.49\linewidth}
  \centering
  \centerline{\includegraphics[width=4.6cm]{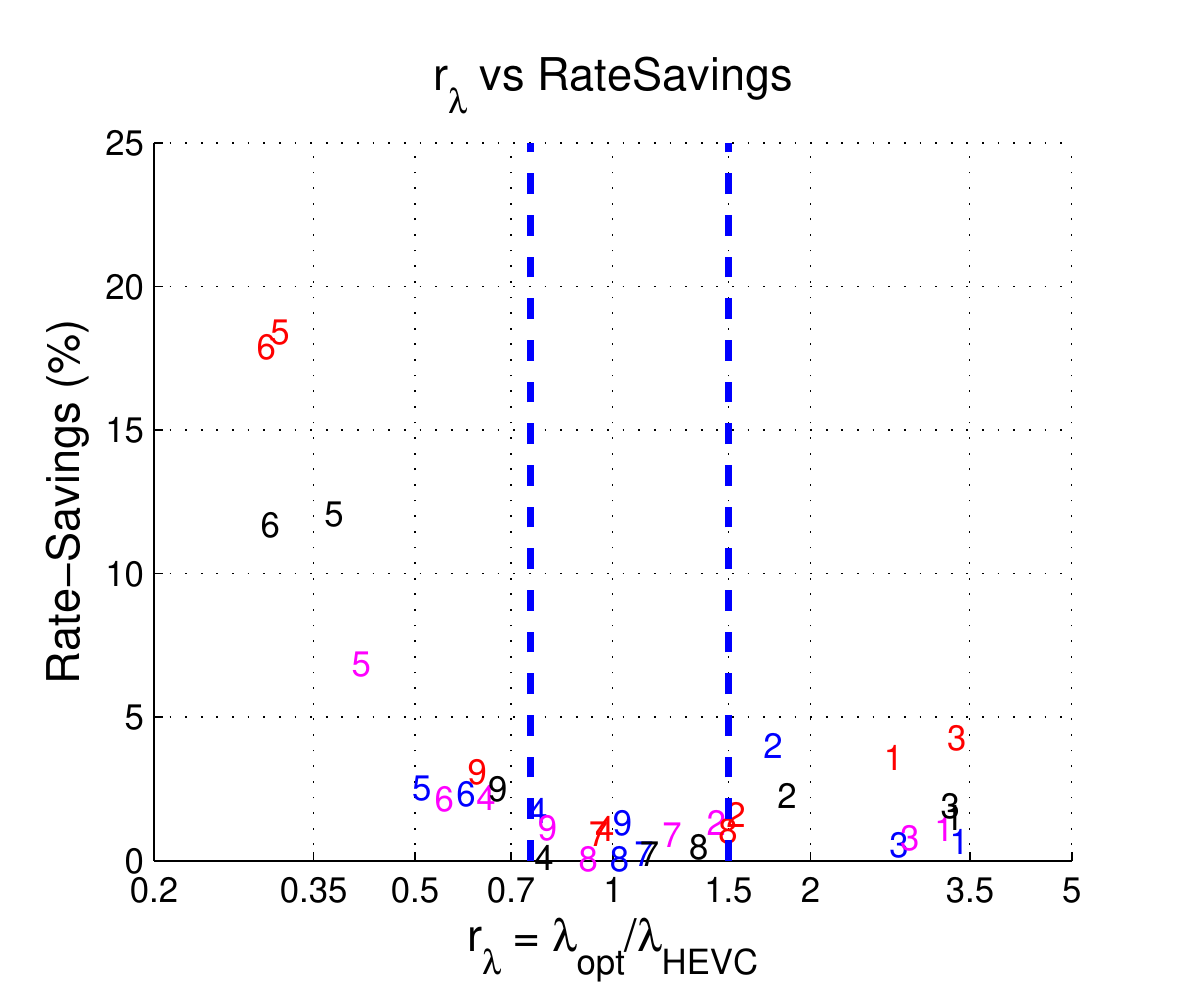}}
\end{minipage} 
\caption{Results for the Lagrange multiplier selection experiment. Row 1: The $\lambda_\mathrm{opt}$ values and $\lambda_\mathrm{orig}$ values versus QP in the H.264/AVC (left) and HEVC (right) codecs. Green curves represent $\lambda_\mathrm{orig}$ as a function of QP, and the position of each number indicates the $\lambda_\mathrm{opt}$ values for that sequence at certain QP. Row 2: The $r_{P/B}$ versus $r_\lambda$ for H.264 (left) and HEVC (right). Row 3: the correspoding bitrate savings for various $r_\lambda$ over H.264 (left) and HEVC (right).}
\label{fig:expresults}
\end{figure}

The optimum Lagrange multipliers ($\lambda_\mathrm{opt}$) are identified for all test sequences and quantisation parameters, based on overall rate-distortion (RD) performance for all frames, benchmarked against the RD performance using $\lambda_\mathrm{orig}$ values. The comparison between $\lambda_\mathrm{opt}$ and their corresponding $\lambda_\mathrm{orig}$ for B frames in both H.264 and HEVC is illustrated in Fig.\ref{fig:expresults} (row 1). It can be seen that the $\lambda_\mathrm{orig}$ values (on the green curve) do not correlate well with $\lambda_\mathrm{opt}$ (points correspond to the numerical values which index the test sequences in Table \ref{tab:testvideos}) for the sequences tested, especially for Class A and B. We believe that for static scene content (Class A), skip mode tends to be frequently invoked which consumes little bitrate, and contravenes the high-rate assumption. For cases with dynamic textures (Class B), intra mode often produces the best results, whereas the original Lagrange multiplier is based on the RD performance using inter prediction.   

This experiment also found an important correlation which can be used to predict optimum Lagrange multipliers for B frames. We firstly define the distortion ratio ($r_{P/B}$) between the average distortion (SSD) of P frames ($D_P$) and that of B frames ($D_B)$ as:
\small
\begin{equation}
r_{P/B}=D_P/D_B
\end{equation}
\normalsize
We also define the ratio between $\lambda_\mathrm{opt}$ and $\lambda_\mathrm{orig}$ (for B frames) as:

\vspace{-10pt}
\small
\begin{equation}
r_\lambda = \frac{\lambda_\mathrm{opt}}{\lambda_\mathrm{orig}}
\end{equation}
\normalsize
The relationship between $r_{P/B}$ and $r_\lambda$ is presented in Fig.\ref{fig:expresults} (row 2) based on the experimental results where distortion statistics are from data using $\lambda_\mathrm{orig}$. 

It can be observed that $r_{P/B}$ correlates well with $r_\lambda$ in most cases. A three parameter power function (green curve in Fig.\ref{fig:expresults} (row 2)) is employed here to fit scatter points, and this is given as follows.
\small
\begin{equation}
r_\lambda = f(r_{P/B} ) = a \cdot r_{P/B}^b+c.
\label{eq:correlation}
\end{equation} 
\normalsize
in which $a$, $b$, $c$ are three constant parameters, and their values are determined based on the results from the nine test sequences, and shown in Table \ref{tab:params}. 

\begin{table}[htbp]
\footnotesize
\caption{Parameters used in both H.264/AVC and HEVC codecs for predicting the optimum Lagrange multiplier values.}
\centering
\begin{tabular}{cccccc}
\toprule
Codec &  $a$  & $b$ & $c$ & $r_1$ &  $r_2$     \\
\midrule
H.264/AVC & 2.696 & 10.06 & 0.367 & 0.81 & 0.93\\
HEVC & 2.197 & 5.196 & 0.308 & 0.73 & 0.89\\
\bottomrule
\end{tabular}
\label{tab:params}
\end{table}

The bitrate savings corresponding to $\lambda_\mathrm{opt}$ compared with the rate-distortion performance using $\lambda_\mathrm{orig}$ are also shown in Fig.\ref{fig:expresults} (row 3). It can be noted that,  when the distortion ratio $r_{P/B}$ drops into a certain band for both codecs (between the two blue dotted bars), the corresponding rate savings decrease significantly.

\section{Proposed Algorithm}
\label{sec:algorithm}

Inspired by the experimental results described above, we propose an in-loop Lagrange multiplier determination method which adaptively predicts optimum Lagrange multipliers based on distortion indices from recent encoded frames. This operates under the assumption that within a temporally localised group of frames, the rate-distortion characteristics are uniform (except for cases where there is  a scene cut, which needs to be detected). The proposed algorithm thus consists of three sub stages: (i) detect a scene cut; (ii) record distortion information; (iii) modify the Lagrange multiplier.

\subsection{Detect a scene cut} 

Scene cut detection, also known as shot transition detection, is a well established research area in video processing. Measures such as sum of absolute differences (SAD), histogram differences (HD) and edge change ratio through static or dynamic thresholding can be used to detect a shot cut \cite{j:Cotsaces}. In our work, in order to reduce the computational complexity, a simple method using histogram differences with a constant threshold is employed. However it should be noted that none of the test sequences reported here contain scene cuts. 
%
%
%

\subsection{Record distortion information} 

To maintain an efficient adaptation of an existing Lagrange multiplier, distortion information from at least one GOP should be continuously recorded. The distortion indices ($D_P$ and $D_B$) for P and B frames are adaptively updated with the mean squared error ($\mathrm{MSE}_t$) of the latest encoded frame using (\ref{eq:update}).

\vspace{-10pt}
\small
\begin{equation}
\left\{ 
\begin{array}{l}
D = 0.2 D + 0.8 \mathrm{MSE}_t, \text{if $D>0$}\\
D = \mathrm{MSE}_t, \text{if $D=0$}\\
\end{array}
\right..
\label{eq:update}
\end{equation} 
\normalsize     
where $D$ could be either $D_P$ or $D_B$ depending on the type of frame. It should be noted that distortion indices are formed from the weighted average of previous indices and the latest distortion, where the latter term is given a higher weighting.

\subsection{Modify the Lagrange multiplier} 

When sufficient distortion information has been recorded, the ratio between distortion indices from B and P frames  $r_{P/B}$ will be calculated, and the Lagrange multiplier is modified based on (\ref{eq:correlation}). According to the analysis in Section \ref{sec:experiment} (Fig. \ref{fig:expresults}), when $r_{P/B}$ falls in a certain band ($r_1$, $r_2$), the bitrate savings with the modified Lagrange multiplier fall below a threshold and become insignificant and even negative. In this case, the Lagrange multiplier for the last B frame is reused for the current frame. This is described as follows:

\vspace{-4pt}
\small   
\begin{equation}
\left\{ 
\begin{array}{l}
\lambda_\mathrm{mdf} = \lambda_\mathrm{last}, \text{if $r_1<r_{P/B}<r_2$}\\
\lambda_\mathrm{mdf} = \lambda_\mathrm{last} (a \cdot  r_{P/B}^b+c), \text{otherwise}\\
\end{array}
\right..
\label{eq:modify}
\end{equation}  
\normalsize
in which, $\lambda_\mathrm{mdf}$ is the modified Lagrange multiplier for the current frame, and $\lambda_\mathrm{last}$ represents the Lagrange multiplier used for the last encoded B frame. The latter can be either the modified or the original. The parameter values for $r_1$ and $r_2$ are shown in Table \ref{tab:params}, based on the experimental results reported in Section \ref{sec:experiment}.

In order to maintain a consistent video quality, the change of Lagrange multiplier between the current and last encoded B frames is constrained ($\pm5\%$), as shown in (\ref{eq:clip}).

\vspace{-10pt}
\small
\begin{equation}
\left\{ 
\begin{array}{l}
\lambda_\mathrm{mdf} = 0.95 \cdot \lambda_\mathrm{last}, \text{if } \lambda_\mathrm{mdf} < 0.95 \cdot \lambda_\mathrm{last}\\
\lambda_\mathrm{mdf} = 1.05 \cdot \lambda_\mathrm{last}, \text{if } \lambda_\mathrm{mdf} > 1.05 \cdot \lambda_\mathrm{last}\\
\end{array}
\right..
\label{eq:clip}
\end{equation}  
\normalsize
The algorithm is then reiterated for all B frames, and the current $\lambda_\mathrm{mdf}$ is used for next frame as $\lambda_\mathrm{last}$.  

\section{Results and Evaluation}
\label{sec:results}

The proposed adaptive Lagrange multiplier selection method has been integrated into the H.264/AVC and HEVC reference models, and compared with both original codecs. Identical test conditions are used here as in the experiment in Section \ref{sec:experiment}, except that uniform test quantisation parameters (QP) are used for all I, P and B frames from QP 22 to QP 42 with an interval of 5 for both codecs. 

\begin{table}[ht]
\centering
\footnotesize
\caption{Test clips used for evaluating the compression performance.}
\begin{tabular}{c|l|l}
\toprule
Group \&  &\multirow{2}{*}{Sequence and  Length}			&\multirow{2}{*}{Source}	  \\
Class &   & \\
\midrule
I.A					& Akiyo (300f), News (300f), Silent (300f) & Standard\\
I.B					& Shadow (300f), Shower (250f), Wheat (250f) & DynTex\\
I.C					& Bus (150f), Tempete (300f), Soccer (300f) & Standard\\
\midrule
II-IV.A					& Clouds (300f), Fungus (300f), Squirrel (300f) & BVI\\
II-IV.B					& Drops (300f), Plasma (240f), Sparkler (300f) & BVI\\
II-IV.C					& Cactus (150f), ParkScene (300f), Tennis (300f) & HEVC\\
\bottomrule			
\end{tabular}
\label{tab:videos}
\end{table}

Four groups of progressive format test sequences (YUV 4:2:0) are used here including nine videos at $352\times288$ resolution (Group I), nine at $416\times240$ (Group II), nine at $832\times480$ (Group III), and nine at $1920\times1080$ (Group IV). These are all public video sequences from the standard test sequence group (I.A and I.B), the HEVC recommended test database \cite{j:Ohm} (II-IV.C), the DynTex database \cite{j:dyntex} (I.B), and the BVI video texture database \cite{w:bviTex} (II-IV.A and II-IV.B). Each group of sequences are further divided into three sub-classes according to the primary video characteristics: static scene (Class A), dynamic scene (Class B) and mixed scene (Class C). It is also noted that Groups II-IV include sequences with the same video content but at different spatial resolutions. This is done in order to investigate the influence of spatial resolution on the proposed method. Table \ref{tab:videos} summarises the grouping of these test sequences, which are all accessible online\fnref{ftn:webdata}. 

The compression performance of the proposed algorithm for both H.264 and HEVC is compared with the corresponding anchor codecs. The results are based on the Bjontegaard delta measurements \cite{j:Bjontegaard} over all frames and are shown in Table \ref{tab:results}. 

\begin{table}[ht]
\centering
\footnotesize
\caption{Summary of the compression results.}
\begin{tabular}{c | c c | c c}
\toprule
 Anchor 					 &    \multicolumn{2}{c|}{H.264 }  		&   \multicolumn{2}{c}{HEVC }	\\
 \midrule
Group \& Class  		&  				BD-PSNR & BD-Rate  						&   BD-PSNR &  BD-Rate         \\
\midrule
I.A					& 0.17dB & -3.9\% & 0.37dB & -7.9\% \\
I.B					& 0.69dB & -11.4\% &	0.26dB & -4.8\% \\
I.C					& 0.00dB & 0.0\% & 0.01dB & -0.5\%\\
\midrule
II.A					& 0.21dB & -4.0\% & 0.14dB & -2.8\% \\
II.B					& 0.51dB & -7.0\% &	0.16dB & -2.9\% \\
II.C					& 0.03dB & -0.7\% & 0.08dB & -2.0\%\\
\midrule
III.A					& 0.20dB & -4.4\% & 0.14dB & -3.4\% \\
III.B					& 0.44dB & -6.0\% &	0.16dB & -2.7\% \\
III.C					& 0.02dB & -0.4\% & 0.09dB & -2.3\%\\
\midrule
IV.A					& 0.16dB & -3.4\% & 0.12dB & -2.9\% \\
IV.B					& 0.44dB & -6.2\% &	0.11dB & -2.1\% \\
IV.C					& 0.03dB & -1.0\% & 0.07dB & -2.4\%\\
\midrule
\midrule
I  						& 0.29dB & -5.1\% & 0.21dB & -4.4\% \\
II  					& 0.25dB & -3.9\% & 0.13dB & -2.6\% \\ 
III  					& 0.22dB & -3.6\% & 0.13dB & -2.8\% \\ 
IV  					& 0.21dB & -3.5\% & 0.10dB & -2.4\% \\  
\midrule
A  						& 0.19dB & -3.9\% & 0.19dB & -4.2\% \\
B  					& 0.52dB & -7.6\% & 0.17dB & -3.1\% \\ 
C 					& 0.02dB & -0.5\% & 0.06dB & -1.8\% \\ 
\midrule
\midrule
Overall         & 0.24dB & -4.0\% & 0.14dB & -3.0\% \\
\bottomrule			
\end{tabular}
\vspace{-8pt}
\label{tab:results}
\end{table}

It can be observed that the proposed method always performs better than both anchor codecs in all test sequence groups and classes, with average bitrate savings of 4\% and 3\% for H.264 and HEVC respectively. It is also noted that this improvement is consistent across various spatial resolutions. Moreover, more significant bitrate savings are seen on static and dynamic scene content (Class A and B) than on video sequences with mixed content (Class C). 

It should also be noted that the proposed algorithm is based on experimental results using uniform QP for all frames - one of the most commonly used scenarios. For cases with different QP values for different frame types, the relationship between B and P frame distortion ratio and optimum Lagrange multipliers may vary. In terms of computational complexity, even when including computation for scene cut detection (linear time $O(n)$ - where $n$ is the number of pixels in each frame), the increased complexity of our approach is negligible in the overall context of H.264/AVC or HEVC compression.  

\section{Conclusion}
\label{sec:conclusion}

A novel adaptive Lagrange multiplier selection method has been presented for rate-distortion optimisation in hybrid video codecs. Based on experimental results, the distortion ratio between B and P frames is employed to modify the Lagrange multipliers used for B frame RDO in uniform QP applications. This method has been integrated into both H.264 and HEVC codecs, and offers a consistent bitrate savings over various test sequences with different spatial resolution and content. Future work will focus on investigating the influence of using varied quantisation parameters, various GOP sizes and accurate Lagrange multiplier determination at block level.
  
\section{REFERENCES}
\label{sec:ref}

\small
\bibliographystyle{IEEEbib}
\bibliography{MyRef}

\begin{thebibliography}{10}

\bibitem{r:cisco}
``Cisco visual networking index: forecast and methodology, 2013-2018,''
\newblock Tech. {R}ep., CISCO, June 2014.

\bibitem{j:HEVC}
G.~J. Sullivan, J.~R. Ohm, W.~J. Han, and T.~Wiegand,
\newblock ``Overview of the high efficiency video coding ({HEVC}) standard,''
\newblock {\em IEEE Transactions on Circuits and Systems for Video Technology},
  vol. 22, no. 12, pp. 1649--1668, 2012.

\bibitem{j:Sullivan1}
G.~Sullivan and T.~Wiegand,
\newblock ``Video cmpression: from concepts to the {H.264/AVC} standard,''
\newblock {\em Proc. IEEE}, vol. 93, no. 1, pp. 61--79, 2005.

\bibitem{j:Sullivan2}
G.~J. Sullivan and T.~Wiegand,
\newblock ``Rate-distortion optimization for video compression,''
\newblock {\em IEEE Signal Processing Magazine}, vol. 15, pp. 74--90, 1998.

\bibitem{b:jayant}
N.~Jayant and P.~Noll,
\newblock {\em Digital Coding of Waveforms},
\newblock Prentice Hall, 1984.

\bibitem{j:Gish}
H.~Gish and J.~N. Pierce,
\newblock ``Asymptotically efficient quantizing,''
\newblock {\em IEEE Transactions on Information Theory}, vol. 14, pp. 676--683,
  1968.

\bibitem{c:Wiegand}
T.~Wiegand and B.~Girod,
\newblock ``{L}agrange multiplier selection in hybrid video coder control,''
\newblock in {\em Proc. IEEE Int Conf. on Image Processing}. IEEE, 2001,
  vol.~3, pp. 542--545.

\bibitem{j:Li}
X.~Li, N.~Oertel, A.~Hutter, and A.~Kaup,
\newblock ``Laplace distribution based {L}agrangian rate distortion
  optimization for hybrid video coding,''
\newblock {\em IEEE Transactions on Circuits and Systems for Video Technology},
  vol. 19, pp. 193--205, 2008.

\bibitem{c:Chen}
L.~Chen and I.~Garbacea,
\newblock ``Adaptive lambda estimation in lagrangian rate-distortion
  optimization for video coding,''
\newblock in {\em Proc. SPIE 6077, Visual Communications and Image Processing}.
  SPIE, 2006.

\bibitem{j:Zhao}
X.~Zhao, J.~Sun, S.~Ma, and W.~Gao,
\newblock ``Novel statistical modeling, analysis and implementation of
  rate-distortion estimation for {H.264/AVC} coders,''
\newblock {\em IEEE Transactions on Circuits and Systems for Video Technology},
  vol. 20, no. 5, pp. 647--660, 2010.

\bibitem{b:Bull}
D.~R. Bull,
\newblock {\em Communicating pictures: a course in image and video coding},
\newblock Academic Press, 2014.

\bibitem{j:dyntex}
R.~P\'eteri, S.~Fazekas, and M.~J. Huiskes,
\newblock ``{DynTex}: a comprehensive database of dynamic textures,''
\newblock {\em Pattern Recognition Letters}, vol. 31, pp. 1627--1632, 2010,
\newblock http://projects.cwi.nl/dyntex/.

\bibitem{w:bviTex}
M.~A. Papadopoulos, F.~Zhang, D.~Agrafiotis, and D.~R. Bull,
\newblock ``{BVI} video texture database,''
  http://data.bris.ac.uk/data/dataset/1if54ya4xpph81fbo1gkpk5kk4.

\bibitem{j:Cotsaces}
C.~Cotsaces, N.~Nikolaidis, and I.~Pitas,
\newblock ``Video shot detection and condensed representation. a review,''
\newblock {\em IEEE Signal Processing Magazine}, vol. 23, no. 2, pp. 28--37,
  2006.

\bibitem{j:Ohm}
J.~R. Ohm, G.~J. Sullivan, H.~Schwarz, T.~K. Tan, and T.~Wiegand,
\newblock ``Comparison of the coding efficiency of video coding standard -
  including {H}igh {E}fficiency {V}ideo {C}oding ({HEVC}),''
\newblock {\em IEEE Transactions on Circuits and Systems for Video Technology},
  vol. 22, no. 12, pp. 1669--1684, 2012.

\end{thebibliography}

\end{document}